\documentclass[aps,twocolumn,prl,groupedaddress]{revtex4}
\usepackage{graphicx}
\usepackage{color}
\usepackage{amsmath}
\usepackage{bm}

\newcommand{\be}{\begin{equation}}
\newcommand{\ee}{\end{equation}}
\newcommand{\bea}{\begin{eqnarray}}
\newcommand{\eea}{\end{eqnarray}}

\renewcommand{\vec}[1]{{\bf #1}}
\renewcommand{\epsilon}{\varepsilon}

\begin{document}

\title{Fermi arc plasmons in Weyl semimetals}
\author{Justin C. W. Song$^{1,2}$ and Mark S. Rudner$^3$}
\affiliation{$^1$Division of Physics and Applied Physics, Nanyang Technological University, Singapore 637371}
\affiliation{$^2$Institute of High Performance Computing, Agency for Science, Technology, and Research, Singapore 138632}
\affiliation{$^3$ Center for Quantum Devices and Niels Bohr International Academy, Niels Bohr Institute, University of Copenhagen, 2100 Copenhagen, Denmark}
\begin{abstract}
In the recently discovered Weyl semimetals, the Fermi surface may feature disjoint, open segments -- the so-called Fermi arcs --  
associated with topological states bound to exposed crystal surfaces. 
Here we show that the collective dynamics of electrons near such surfaces sharply departs from that of a conventional three-dimensional metal. 
In magnetic systems with broken time reversal symmetry, the resulting Fermi arc plasmons (FAPs) are chiral, with dispersion relations featuring open, hyperbolic constant frequency contours. 
As a result, a large range of surface plasmon wave vectors can be supported at a given frequency, with corresponding group velocity vectors directed along a few specific  
collimated directions. Fermi arc plasmons can be probed using near field photonics techniques, which may be used to launch highly directional, focused surface plasmon beams. 
The unusual characteristics of FAPs arise from the interplay of bulk and surface Fermi arc carrier dynamics, and give a window into the unusual Fermiology of Weyl semimetals.
\end{abstract} 
%\pacs{}

\maketitle

\begin{figure}[t]
\includegraphics[width=\columnwidth]{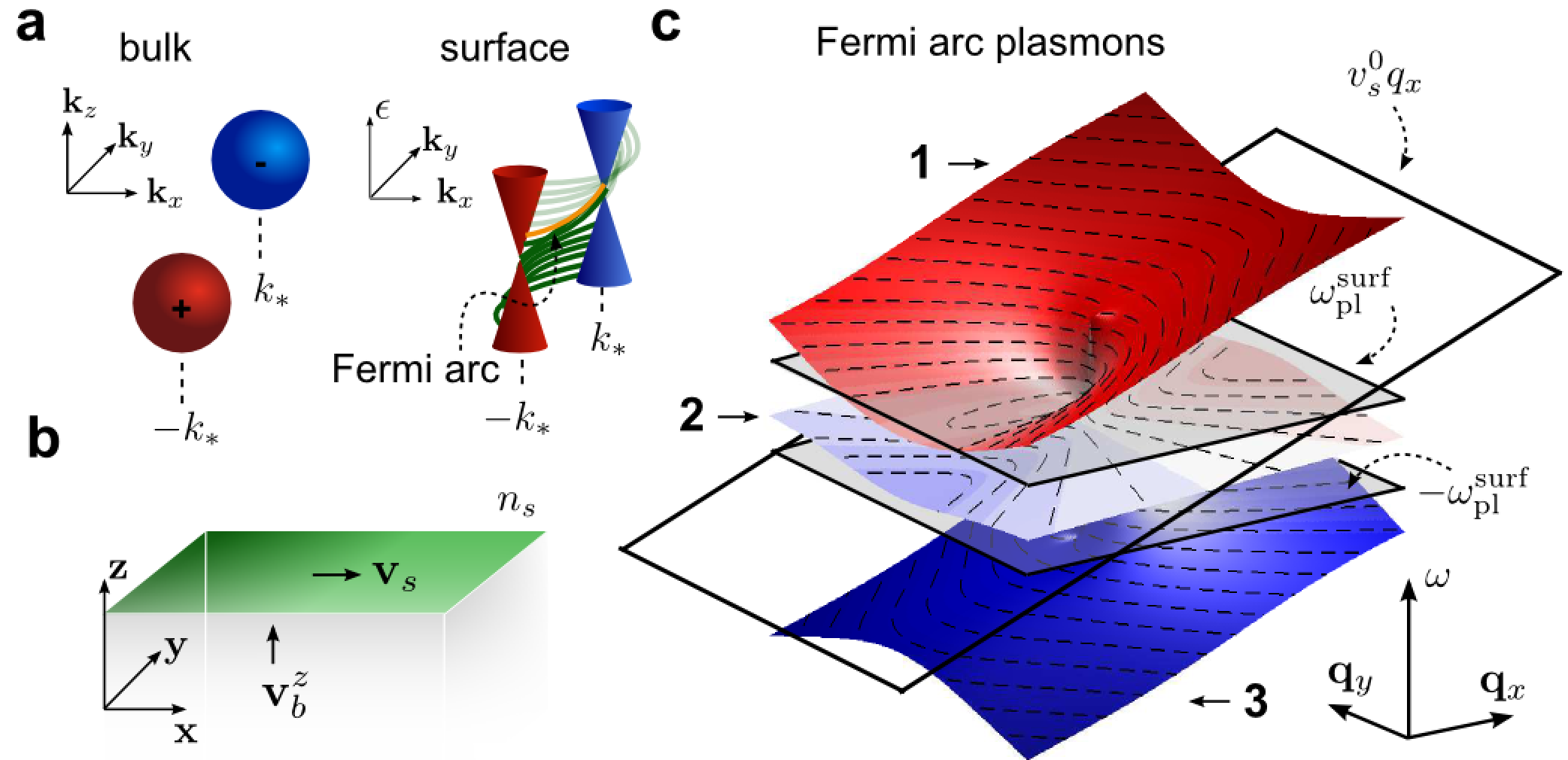}
\caption{{\bf a}. Disjoint Fermi surface of a type-I Weyl semimetal: in addition to the bulk 
closed Fermi surfaces (red and blue spheres), the WSM features open Fermi arcs (orange curve) on  
certain exposed crystal surfaces. 
{\bf b}. The velocity density can be separated into bulk ($\vec v_b$) and surface ($\vec v_s$) components, with $\vec v_s$ 
carried by the chiral Fermi arc states. 
{\bf c}. The Fermi arc plasmon dispersion has three chiral branches, featuring hyperbolic frequency contours over a wide bandwidth.
These modes arise from the hybridization of conventional surface plasmons, supported by bulk free carriers, and collective modes of carriers in the topological surface states.
The ordinary surface plasmon frequency is denoted by $\pm\omega_{\rm pl}^{\rm surf}$, and topological surface states are characterized by the velocity $v_s^0$.}\label{fig:summary}
\end{figure}

Three dimensional Weyl semimetals (WSMs) are prototypical topological metals, featuring one or more pairs of protected band degeneracy points in the Brillouin zone~\cite{yingran,balents,wan,xugang,ding15,hasan15, TurnerVishwanath}.
These ``Weyl points'' act as quantized sources and sinks of Bloch band Berry flux~\cite{Haldane2004}.
The nontrivial bulk band topology is manifested in the appearance of peculiar branches of gapless states bound to certain exposed surfaces, featuring {\it open segment} Fermi surfaces, i.e., ``Fermi arcs''(see Fig.~\ref{fig:summary}a)~\cite{yingran,balents,wan,xugang,ding15,hasan15, HaldanePlumbing}.
At the single particle level, these unusual features are responsible for a variety of intriguing transport phenomena~\cite{potter,analytis,ong,sonspivak,spivakandreev,xidai, baum,Sid}.

\begin{figure}[t]
\includegraphics[width=\columnwidth]{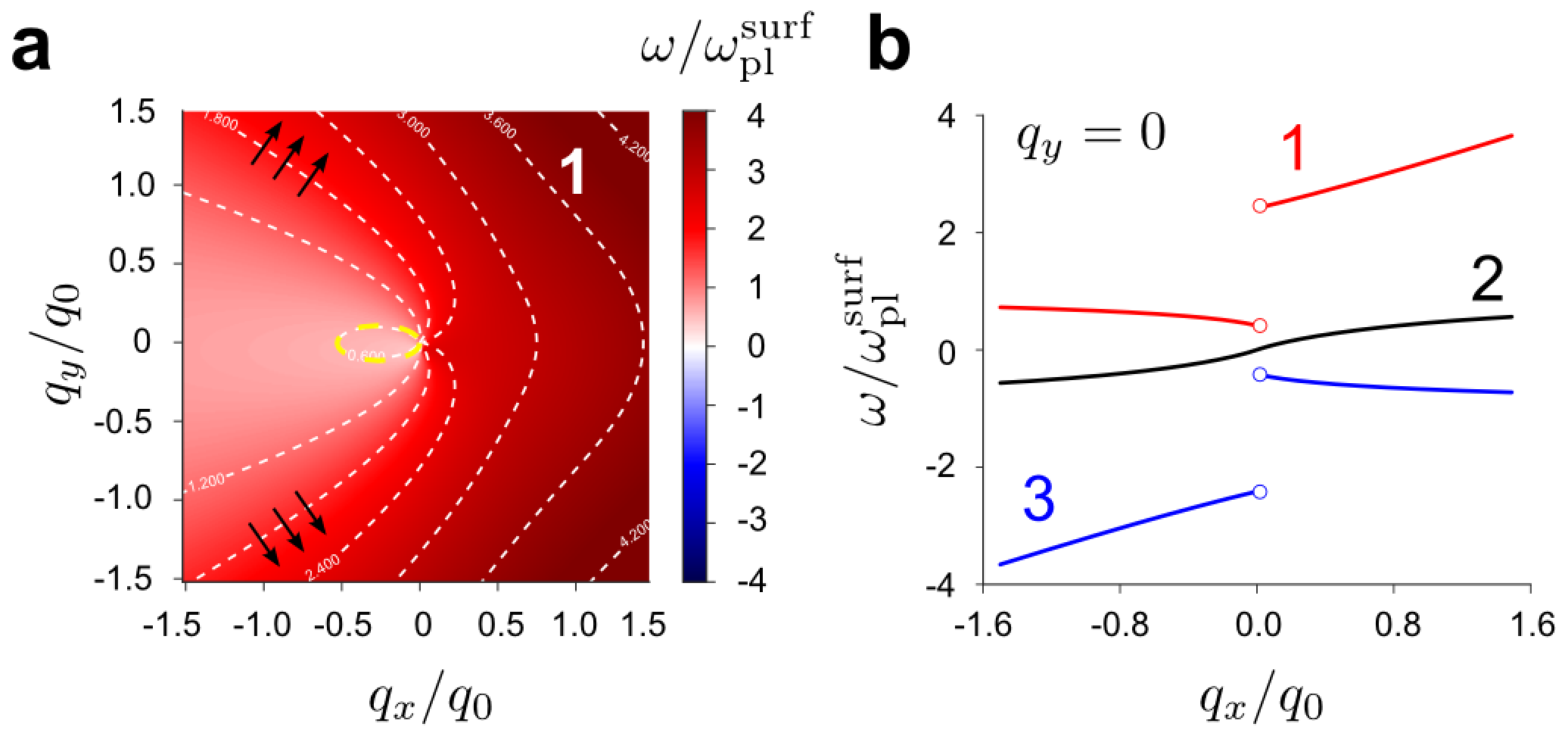}
\caption{{\bf a}. Contour plots of the FAP dispersion for branch 1 
in Fig.~\ref{fig:summary}c, obtained from Eq.~(\ref{eq:decoupledsecular}). 
Wave vectors are scaled by $q_0=\omega_{\rm pl}/v_s^{0}$.
Fine dashed contours are hyperbolic, and do not close on themselves at large wave vectors.
In a small frequency interval near $|\omega| \lesssim \omega_{\rm pl}^{\rm surf}$, elliptical contours are found (bold yellow dashed lines, see discussion in text).
The FAP group velocity is oriented transverse to the constant frequency contours (black arrows).
Along a given hyperbolic frequency contour, the group velocity points along a single direction for a large range of $q_y$ values, allowing for focused propagation of plasmon waves.
{\bf b}. Line cut of FAP dispersion for $q_y=0$. 
Parameters used: dimensionless Drude weight $\tilde{\mathcal{D}} = 6.0$, see Eq.~(\ref{eq:asymptotic}), and dimensionless Hall conductivity $\tilde{\sigma}_H =4\pi \sigma_H/[(\kappa +1) \omega_{\rm pl}^{\rm surf} ]=2.0$~\cite{parameters}.
}
\label{fig:dispersion}
\end{figure}

Here we show that the coupled, collective dynamics of electrons on both the open-segment and closed WSM Fermi surfaces sharply departs from that expected in conventional metals. 
This unusual situation gives rise to ``Fermi arc plasmons'' (FAPs), which are confined to certain exposed surfaces and are characterized by highly anisotropic dispersions
(Figs.~\ref{fig:summary}b,c).  
In conventional metals, surface plasmons are nearly dispersionless and feature closed, elliptical constant frequency contours.
In contrast, we find that FAPs in WSMs with broken time reversal symmetry
are 
{\it hyperbolic} over a wide range of frequencies, featuring open 
iso-frequency contours that do not close on themselves (Fig.~\ref{fig:dispersion}). 
Such 
contours support a wide range of wave vectors at each frequency, with a wavenumber-independent group velocity direction for large wave vectors. 
Together, these features allow for tight focusing of 
collimated, non-reciprocal surface plasmon waves, with frequency-dependent directionality.

%%%%%%%%%%%%%%%%%%%%%%%%%%%%%%%%%%%%%%%%%%%%%%%%%%%%%%%%%%%
%%%%%%%%%%%%%%%%%%%%%%%%%%%%%%%%%%%%%%%%%%%%%%%%%%%%%%%%%%%

FAPs arise from the hybridization of chiral collective modes associated with carriers in topological ``Fermi arc'' surface states (slanted plane, Fig.~\ref{fig:summary}c), 
with ordinary surface plasmon modes supported by bulk carriers close to the surface (horizontal gray planes, Fig.~\ref{fig:summary}c).
This hybridization and the resulting dispersion relations 
are sensitive to the unusual constitutive relations of the Weyl semimetal's bulk and surface carriers, including the bulk  
anomalous Hall conductivity~\cite{burkov}, and the chiral Fermi arc surface state velocity~\cite{wan,yingran,balents}. 
Consequently, FAPs  
may also provide means to dynamically probe the peculiar carrier kinematics of Weyl semimetals at zero-field~\cite{burkov,son}.  

We expect that FAPs can be realized in WSMs that break time reversal symmetry (TRS), such as candidate magnetic WSMs~\cite{sushkov,borisenko,yingran,wan,balents,tong,bulmash,bernevig16}.
We note parenthetically that hyperbolic plasmons can be artificially engineered in metamaterials, where hyperbolic characteristics are manifested for plasmon wavelengths larger than the feature sizes of resonator components~\cite{hyperbolicrev}. 
In contrast, WSMs that host FAPs are {\it intrinsic} hyperbolic plasmonic materials, and may exhibit hyperbolic behavior at 
short wavelengths, not limited by fabrication or patterning techniques. 
Indeed, FAPs may arise for plasmon wave vectors up to 
the characteristic Weyl node separation, $k_*$, which is typically an $\mathcal{O}(1)$ fraction of the inverse lattice constant: 
$k_* \sim 1 \, {\rm nm}^{-1}$. 
This scale 
corresponds to wavelengths much smaller than typical patterned feature sizes, which may be on the order of $\gtrsim 100\, {\rm nm}$~\cite{hyperbolicrev}. 

Previous studies have revealed a number of interesting plasmonic phenomena, including non-reciprocity~\cite{hoffman16} and signatures of the chiral anomaly~\cite{sonspivak,pesin,xiao,spivakandreev,hoffman16} 
associated with the collective behavior of carriers on the (closed) bulk Fermi surfaces of WSMs~\cite{Rosenstein, Cortijo, Lozovik}.
Here we aim to expose the new collective phenomena and coupling mechanisms that arise due to the chiral Fermi arc modes on the WSM surface.

\vspace{2mm}
{\bf Collective carrier dynamics} --- 
To begin, we
examine collective charge carrier dynamics 
in a 3D WSM in 
a semi-infinite slab geometry, occupying the region $z < 0$, see Fig.~\ref{fig:summary}b.
For concreteness, we consider a model dispersion with two bulk Weyl nodes at zero energy, situated at points $(0, \pm k_*, 0)$ in the Brillouin zone (i.e., separated along the $k_y$ direction)~\cite{SI}. 
For each fixed value of $k_y$ with $|k_y| < k_*$, the two-dimensional dispersion $\epsilon_{2D}(k_x, k_z; k_y)$ is gapped, with 
upper and lower bands featuring Chern numbers +1 and -1, respectively. At the surface $z = 0$, this nontrivial topology yields a branch of gapless chiral surface states for each $|k_y| < k_*$. 
When the Fermi energy is close to zero, 
the filling of the chiral surface states is characterized by an {\it open segment} Fermi surface -- the ``Fermi arc'' (see Fig.~\ref{fig:summary}a, and Refs.~\cite{wan,yingran,balents}).
As we show, these Fermi arc carriers bring dramatic new features to 
WSM surface plasmons. 

To describe FAPs in the slab geometry, it is useful to partition 
the particle density field $n$ and the velocity density field $\vec{v}$ into 
bulk and surface contributions.
For each Fourier mode with
angular frequency $\omega$, 
we write:
\bea
\nonumber n(\vec r, \omega) &=&  n_b(\vec r, \omega) \Theta(-z) + n_s(\vec r_s, \omega) \delta(z), \\  \vec v (\vec r, \omega) &=& \vec v_b(\vec r, \omega) \Theta (-z) + \vec v_s(\vec r_s, \omega) \delta (z). 
\label{eq:separate}
\eea 
Here the $s$ and $b$ subscripts of $n_{s,b}$ and $\vec{v}_{s,b}$ denote surface and bulk contributions, respectively, and $\vec{r}_s = (x, y)$ is the two-dimensional coordinate on the surface $z = 0$. The surface densities $n_s$ and $\vec{v}_s$ account both for surface charges that may build up from the motion of bulk free carriers, as in a usual metal~\cite{fetter-b}, and for the behavior of carriers in the unconventional topological surface
states. 
We consider the case of low but finite doping, so that $k_*$ is large as compared to the typical Fermi wave vectors in the bulk.
Note that $\vec{v}_s$ is always oriented within the plane of the two-dimensional surface, so that $v_s^z = 0$.

The bulk carrier density obeys the continuity relation
 \be
  i\omega n_b  + \nabla \cdot \vec v_b  = 0,\ \  e {\vec{v}}_b (\vec{r}, \omega) = \bm{\sigma}(\vec{r}, \omega) \big[-\nabla \phi(\vec{r}, \omega)\big], 
 \label{eq:currentcon}
\ee
where $\bm{\sigma}(\vec{r}, \omega) \propto \Theta(-z)$ is the bulk conductivity tensor, 
$\phi$ is the electric potential, and $e = -|e|$ is the electron charge. The conductivity $\bm{\sigma}(\omega)$ includes an intrinsic anomalous Hall component $\sigma_{H} = \sigma_{zx} = -\sigma_{xz}$, due to the Berry flux between the Weyl nodes~\cite{burkov}, and a longitudinal part $\sigma_{xx} = \sigma_{yy} = \sigma_{zz}$. Note that we work at zero applied magnetic field, and do not access the finite magnetic field WSM collective dynamics~\cite{sonspivak,pesin,baum,xiao,spivakandreev,hoffman16}.

At the surface $z = 0$, the continuity relation for the surface density picks up a source term arising from bulk currents impinging on the surface: 
\be
i\omega n_s(\vec r_s, \omega)  +  \nabla \cdot \vec v_s (\vec r_s, \omega) =  \hat{\vec{z}} \cdot \vec v_b \big|_{0_-}.
\label{eq:surface}
\ee 
To arrive at Eq.~(\ref{eq:surface}), we have used Eq.~(\ref{eq:separate}) and the full continuity equation, along with $\partial_z \Theta (-z) = - \delta (z)$.
Plasmons emerge from Eqs.~(\ref{eq:currentcon}) and (\ref{eq:surface}) as self-sustained collective oscillations, with the electric potential 
generated internally by density fluctuations $\delta n (\vec r, \omega)= n(\vec r, \omega) - n_0$. 
Here $n_0$ is the equilibrium electron density. 

As in a conventional metal, 
bulk carriers (denoted $M$) may pile up at a 
surface when the system is pushed out of equilibrium, giving a surface density contribution $n_s = n_s^{M}$~\footnote{Here we neglect the role of incidental, trivial surface states that may be present for certain surface terminations.}.
In equilibrium, $n_s^M \equiv 0$.
Importantly, 
the associated surface velocity density, $\vec v^M_s$, 
vanishes at linear order in the deviation from equilibrium~\cite{SI}. 
In the linear regime, Eq.~(\ref{eq:surface}) together with 
the non-retarded Coulomb interaction  yields the familiar non-dispersive surface plasmon mode of a 3D metal~\cite{ritchie}.

For the WSM, in addition to the conventional surface density $n_s^M$, we must also account for the dynamics of carriers in the topological surface states (denoted $F$). 
We thus express the total surface density as $n_s(\vec{r}_s, t) = n_s^M(\vec{r}_s, t) + n_s^F(\vec{r}_s, t)$~\footnote{While topologically trivial surface states may also exist, they do not give rise to the peculiar FAP phenomenology that we unveil.}, where
$n_s^F(\vec{r}_s, t)$ is the Fermi arc surface state density distribution (see Figs.~\ref{fig:summary}a,b).
Here $n_s^F$ includes all electrons in the topological surface states up to the Fermi arc, and is thus sensitive to details of the dispersion far below the Fermi surface~\footnote{A natural cut-off for the topological surface state bandwidth, and hence for $n_s^{0}$, can be estimated from the energy of the Lifshitz point where the Weyl cones touch.}. However, the electrodynamic response only depends on deviations from equilibrium, i.e., deformations of the Fermi arc.

The topological surface states carry a finite (in-plane) surface velocity density, analogous to the persistent currents that flow in quantum Hall edge states. For a straight Fermi arc with $k_y$-independent dispersion, $\epsilon_{s}(k_x, k_y) = v_s^0 k_x$, the surface velocity density is proportional to $n_s^F$ and oriented along $\hat{\vec{x}}$. Here $v_s^0$ is the velocity of the chiral surface states on the surface $z = 0$; it has the same sign as $\sigma_H$~\cite{SI}. For a more general topological surface state dispersion, non-equilibrium deformations of the Fermi arc may also lead to currents {\it parallel} to $\hat{\vec{y}}$, 
similar to a conventional 2D system with a closed Fermi surface.
We account for 
Drude-like 
dynamics along $\hat{\vec{y}}$ using a phenomenological Drude weight $\mathcal{D}$~\cite{SI} 
and write: 
\be
\vec v_s (\vec r_s, t)= \big[v_s^0 n_s^F (\vec r_s, t)\big] \hat{\vec{x} } + \left[\frac{\mathcal{D}}{i e \omega} \big[-\partial_y \phi(\vec r_s, t)\big]\right] \hat{{\vec y}}. 
\label{eq:vel}
\ee
Note that $\mathcal{D} = 0$ for a system with $k_y$-independent surface state dispersion (as described above).

We now analyze the surface density dynamics, Eq.~(\ref{eq:surface}).
We focus on the collisionless limit, $\omega \gg 1/\tau_{\rm surf}$, where $\tau_{\rm surf}$ is a relaxation time characterizing the
scattering between $F$ and $M$ species on the surface~\footnote{At low temperatures, $1/\tau_{\rm surf} \to 0$, while the oscillation frequency $\omega$ remains finite.}. 
In this limit, $n_s^F$ and $n_s^M$ 
obey separate continuity equations: 
\bea
\nonumber &ie\omega n_s^M + \sigma_{xx} \nabla_z \phi \big|_{0^-}=0\\
&ie\omega n_s^F+  \sigma_{zx}\nabla_x \phi+ e\nabla \cdot \vec v_s = 0.
\label{eq:eom}
\eea
The two fluids $M$ and $F$ are coupled by the Coulomb interaction, encoded 
in $\phi (\vec r, \omega)$.
In the linear regime, $\vec v_s$ is associated solely with the $F$ carriers (see above).

Notice that the surface density dynamics is explicitly coupled to bulk currents in Eq.~(\ref{eq:eom}). 
In particular,  
$\sigma_{xx}\nabla_z\phi\big|_{0^-}$ 
captures the current density of bulk free carriers impinging on the metal surface from $z < 0$.
Similarly, $\sigma_{zx} \nabla_x \phi$ accounts for the fact that an {\it in-plane} electric field induces changes in the topological surface state density, via anomalous Hall currents associated with undergap carriers at momenta far from the Weyl nodes.

\vspace{2mm}
{\bf Fermi arc plasmons in a Weyl semimetal} --- 
FAPs arise as surface-bound plane wave solutions to the equations of motion (\ref{eq:separate})-(\ref{eq:eom}), of the form $\delta n, \phi \sim e^{i\omega t-i \vec{q}\cdot \vec{r}_s}$.  
Here $\vec{q} = (q_x, q_y)$. 
Below we algebraically eliminate the particle and velocity density fields from the equations of motion, obtaining a compact description of the coupled motion of surface and bulk carriers in terms of $\phi(\vec r,\omega)$.

Away from the surface at $z = 0$, 
the Fourier modes $\tilde{\phi}_{\vec{q}}(z)$ inside ($<$) and outside ($>$) the WSM obey the Poisson equation (in the non-retarded limit): 
\be
(q^2 - \partial_z^2) \tilde{\phi}_{\vec{q}}^<(z) =  \frac{4\pi e}{\kappa} \tilde{n}_{b, \vec{q}}(z),\ \ (q^2 - \partial_z^2) \tilde{\phi}_{\vec{q}}^>(z) =0,
\label{eq:potential}
\ee
where $q = |\vec{q}|$, and $\kappa$ is the dielectric constant in the WSM (the region denoted $>$ is vacuum); the $\omega$ dependencies of $\tilde{\phi}_{\vec{q}}$ and $\tilde{n}_{b}(z)$ are implicit. We eliminate $\tilde{n}_{b,\vec{q}}$ in Eq.~(\ref{eq:potential}) using Eq.~(\ref{eq:currentcon}), $i\omega {n}_b = (1/e)\nabla\cdot({\bm \sigma}(\omega)\nabla\phi)$~\footnote{Note that only the symmetric part of $\bm{\sigma}$ contributes to the divergence.}, and obtain solutions $
\tilde{\phi}_{\vec{q}}^<(z) = \alpha_{\vec{q}} e^{qz}, \, \tilde{\phi}_{\vec{q}}^> (z) = \beta_{\vec{q}} e^{-qz}.
\label{eq:phi}
$

To fully determine the plasmon potential $\tilde{\phi}_{\vec{q}}(z)$, we must specify appropriate boundary conditions at the surface $z =0$.
First, $\tilde{\phi}_{\vec{q}}(z)$ must be continuous at $z=0$. 
Second, note that FAPs involve a dynamical 
modulation of the surface charge density, $en_s$. 
As a result, the electric displacement field across the surface exhibits a jump:
\be
\partial_z \tilde{\phi}_{\vec{q}}\big|_{0^+}  -  \kappa\partial_z \tilde{\phi}_{\vec{q}} \big|_{0^-} =  -4\pi e \delta \tilde{n}_{s,\vec{q}}. 
\label{eq:vbc}
\ee 
Here we use $\delta n_s  = \delta n_s^M + \delta n_s^F$, with $n^\chi_s(\vec{r}_s, \omega) = n_s^{\chi, (0)} + \delta n_s^{\chi}(\vec{r}_s, \omega)$, where $\chi = \{M, F\}$.
Superscript $(0)$ denotes the equilibrium carrier densities. Note that there is no jump in the displacement field in equilibrium.

To obtain a closed set of equations for the potential $\tilde\phi_{\vec{q}}(z)$, we must eliminate $\delta \tilde{n}_{s,\vec{q}}$ from Eq.~(\ref{eq:vbc}). Using Eq.~(\ref{eq:eom}) with Eq.~(\ref{eq:vel}) for a plane wave, we relate the surface density fluctuations to the electric potential $\tilde \phi_{\vec{q}}^<$:
\be
\delta \tilde n_{s,\vec{q}}^{\chi} = \mathcal{G}^{\chi} \tilde \phi_{\vec{q}}^{<}, \ \mathcal{G}^M=  - \frac{q \sigma_{xx}}{i e\omega}, \  \mathcal{G}^F =  \frac{q_x \sigma_{zx} + \tfrac{\mathcal{D}}{\omega} q_y^2 }{ e( \omega - v_s^0 q_x)}.
\label{eq:ns}
\ee
Note that the chirality of the Fermi arc carriers is exhibited in Eq.~(\ref{eq:ns}) through the pole in $\mathcal{G}^F$, 
which arises for a {\it single} value of $q_x$ for a given $\omega$.

Using Eq.~(\ref{eq:ns}) in Eq.~(\ref{eq:vbc}), and the explicit forms for $\tilde\phi_{\vec{q}}^>(z)$ and  $\tilde\phi_{\vec{q}}^<(z)$, we seek the values of $\vec{q}$ and $\omega$ such that the boundary conditions on $\tilde\phi_{\vec{q}}$ and $\partial_z\tilde\phi_{\vec{q}}$ are satisfied. This yields the FAP dispersion relation: 
\be
(\omega - v_s^0 q_x)\Big[\frac{\kappa+1}{\kappa} - \frac{\omega_{\rm pl}^2}{\omega^2}\Big] - \frac{4\pi}{\kappa} \frac{q_x}{q}\sigma_H - \frac{4\pi \mathcal{D} q_y^2}{\kappa \omega q}=0.
\label{eq:decoupledsecular}
\ee
Here we have 
used 
 $4\pi\sigma_{xx}/\kappa = {\omega}_{\rm pl}^2/i\omega$, valid in the collisionless limit; $\omega_{\rm pl}$ is the bulk plasmon frequency.

%%%%%%%%%%%%%%%%%%%%%%%%%%%%%%%%%%%%%%%%%%%%%%%%%%%%%%%%%%%
%%%%%%%%%%%%%%%%%%%%%%%%%%%%%%%%%%%%%%%%%%%%%%%%%%%%%%%%%%%

\vspace{2mm}
{\bf Characteristics of Fermi arc plasmons} --- The roots of Eq.~(\ref{eq:decoupledsecular}) yield the frequencies of Fermi arc plasmons. We find three distinct branches of solutions (labeled 1, 2, and 3), as shown in Fig.~\ref{fig:summary}c and Fig.~\ref{fig:dispersion}. 
These branches arise from the hybridization of conventional dispersionless surface plasmons with the collective excitations in the topological surface states. 
The hybridization is controlled by $\sigma_H$ and $\mathcal{D}$, which couple the bulk and surface densities [see Eqs.~(\ref{eq:vel}) and (\ref{eq:eom})]. 
We note that branch 3 is related to branch 1 via the transformation $\omega \to - \omega$ and $\vec q \to -\vec q$.
However, at each $\vec q$ there are three distinct solutions, associated with modes with different velocities.

In Fig.~\ref{fig:dispersion}a we show a contour plot of 
the dispersion relation of the FAP branch labeled 1 in Fig.~\ref{fig:summary}c. 
Interestingly, most of these contours are {\it open}, and resemble hyperbolae: at large $|\vec q|$, the open contours asymptotically approach the line $\theta = \theta_\infty(\tilde\omega)$, with 
\be
\frac{\sin^2\theta_\infty}{\cos\theta_\infty}= - \frac{\tilde{\omega}}{\tilde{\mathcal{D}}} \Big( 1- \frac{1}{\tilde{\omega}^2}\Big),\quad \tilde{\mathcal{D}} = \frac{4\pi \mathcal{D}}{(\kappa+1) v_s^0 \omega_{\rm pl}^{\rm surf}},
\label{eq:asymptotic}
\ee
where $\vec{q} = q\,(\cos\theta, \sin\theta)$,  $ \tilde\omega = \omega/\omega_{\rm pl}^{\rm surf}$, and $ \omega_{\rm pl}^{\rm surf}= \omega_{\rm pl}\sqrt{\kappa/(\kappa+1)}$. 
Eq.~(\ref{eq:asymptotic}) was obtained by taking the large $|\vec q|$ limit of Eq.~(\ref{eq:decoupledsecular}). 
Since $\sin^2\theta_\infty/\cos\theta_\infty$ 
is even under $\theta_\infty \to - \theta_\infty$, there are two asymptotes, see 
Fig.~\ref{fig:dispersion}. 

The FAP velocity on a given branch $j$, and at a given wave vector $\vec{q}$, $d\omega^{(j)}/d\vec q$, is directed perpendicular to the corresponding constant frequency contour at $\vec{q}$ (Fig.~\ref{fig:dispersion}a, black arrows). 
Due to their hyperbolic character,
large wave vector FAPs at a given frequency in Eq.~(\ref{eq:asymptotic}) all propagate along the specific direction $\hat{\vec{v}}_{\rm pl} =  |\sin\theta_{\infty}|\, \hat{\vec{x}} - {\rm sgn}[\sin\theta_{\infty}]\cos \theta_\infty\, \hat{\vec{y}}$, independent of $|\vec q|$ (for $|\vec q| \lesssim k_*$ where our treatment is valid).
Consequently, these FAPs propagate as collimated beams. Further, as allowed by broken TRS, hyperbolic FAP modes are non-reciprocal: as shown in Fig.~\ref{fig:dispersion}, the FAP velocity is predominantly directed towards $+\hat{\vec{x}}$. 

The asymptotic pitch $\theta_\infty$ of each hyperbolic constant frequency contour, and hence the direction $\hat{\vec{v}}_{\rm pl}$ of collimated FAP propagation, is controlled by the frequency $\omega$, see Eq.~(\ref{eq:asymptotic}).  Approaching the conventional surface plasmon frequency $\omega^{\rm surf}_{\rm pl}$ from above, the RHS of Eq.~(\ref{eq:asymptotic}) approaches zero from below and the linear asymptotes bend towards $-\hat{\vec{x}}$: $\theta_\infty \to \pm \pi$. 
Consequently, the FAP velocity $\hat{\vec{v}}_{\rm pl}$ cants 
towards $\hat{\vec{y}}$, the direction perpendicular to the chiral surface state velocity $v_s^0\,\hat{\vec{x}}$, see Eq.~(\ref{eq:vel}). 
This 
behavior for $\omega \gtrsim \omega^{\rm surf}_{\rm pl}$ results from the resonant hybridization of 
intra-Fermi-arc oscillations with the conventional metallic surface plasmon mode. 

A further striking signature of the coupled bulk and surface carrier dynamics of the WSM can be found in the FAP dispersion for  $q_y = 0$ (see Fig.~\ref{fig:dispersion}b), 
where the Drude weight $\mathcal{D}$ drops out of Eq.~(\ref{eq:decoupledsecular}). 
Analyzing the limits $q_x \to 0^\pm$ in Eq.~(\ref{eq:decoupledsecular}), we find a 
discontinuity in the FAP dispersion (within branch 1) between $q_x \to 0^+$ and $q_x \to 0^-$:
\be
\omega_\pm^{(1)} = \sqrt{\big[2\pi \sigma_H/(\kappa+1)\big]^2 + [\omega_{\rm pl}^{\rm surf}]^2} \pm \frac{2\pi \sigma_H}{(\kappa+1)},
\label{eq:split}
\ee
where $\omega_\pm^{(1)} = \omega^{(1)} (q_x \to 0^\pm, q_y=0)$. 
A similar splitting also arises for branch 3. 
However, there is no splitting for branch 2 since $\omega^{(2)} (q_x=0, q_y=0) = 0$. 

The discontinuity in Eq.~(\ref{eq:split}) arises from 
bulk anomalous Hall currents directed into or out-of the surface. 
These currents are driven by the in-plane electric field associated with FAP waves $e^{i\omega t - i q_x x}$ that propagate along $+\hat{\vec{x}}$ $(q_x > 0)$ or $-\hat{\vec{x}}$ $(q_x < 0)$, respectively~\cite{zyuzin,song16,hoffman16}. 
We note that retardation effects may alter the jump discontinuity at very small $q_x$ via hybridization with the light-cone~\cite{hoffman16}.
The splitting $\omega_+^{(1)} - \omega^{(1)}_-$ is proportional to $\sigma_H$, and thus provides 
the basis for a dynamical measurement of the anomalous Hall conductivity in WSMs. 
Interestingly, for small bulk densities, the role of these anomalous Hall currents becomes dominant, and the FAP frequency becomes insensitive to $\omega_{\rm pl}$: $\omega_+^{(1)} \to 4\pi \sigma_H/(\kappa+1)$.

In addition to the hyperbolicity discussed above, FAPs with elliptic constant frequency contours also occur in small frequency windows near $\pm\omega_{\rm pl}^{\rm surf}$, see yellow dashed lines in Fig.~\ref{fig:dispersion}a. Focusing on the case $\omega > 0$ (branch 1), closed contours exist at frequency $\omega$ if there is a finite, positive value of $q$ that solves Eq.~(\ref{eq:decoupledsecular}) for $q_y = 0, q_x < 0$. This condition is realized for $\omega^{(1)}_- < \omega < \omega^{\rm surf}_{\rm pl}$. Similar considerations hold for branch 3.

\vspace{2mm}
{\bf Discussion ---} 
In this work, our aim has been to expose the unique phenomenology of surface plasmons arising from the coupled dynamics of bulk and topological surface state carriers in Weyl semimetals. The hyperbolic character of FAPs can be probed, e.g., via  
scanning near-field optical microscopy (SNOM)~\cite{koppens2012,basov2012}. 
On the surface of a WSM, 
the near-field tip may launch a pair of collimated beams 
along the two directions of $\hat{\vec{v}}_{\rm pl}$ 
consistent with the applied excitation frequency; the propagation is biased along the direction of the chiral velocity, $v_s^0 \hat{\vec{x}}$.  
In contrast, in hyperbolic materials with TRS, collimated plasmon beams are launched in {\it fours}~\cite{alu,low}, with reflection symmetry in $x$ and $y$.

We note that, even without topological surface states, the presence of a bulk anomalous Hall conductivity in a magnetic system can itself 
yield chiral surface or edge plasmons~\cite{zyuzin, song16,hoffman16}. Additionally,
FAPs exhibit hyperbolicity out to large frequencies (eventually 
limited by the topological surface state bandwidth), far beyond that of $\omega_{\rm pl}^{\rm surf}$ (see, e.g., branch 1 in Fig.~\ref{fig:summary}c). 
Thus, directed plasmon beams with a large bandwidth, characteristic of hyperbolic FAPs, 
provide an optical signature of WSM Fermi arc surface states~\cite{wan}.

Before closing, we briefly outline the material characteristics that favor FAP observation. 
For example, plasmon excitations with energies in the bulk interband particle-hole continuum are damped by electron-hole pair creation. 
Hence it is desirable to have a small surface plasmon frequency $\omega_{\rm pl}^{\rm surf}$ (around which the FAP characteristics are most pronounced), which is below the interband threshold. 
Additionally, since $\omega_{\rm pl}^{\rm surf}$ is directly determined by total carrier density, whereas the interband pair creation threshold is determined by the Fermi energy relative to the Weyl nodes, WSMs with low carrier densities, but with moderately high Fermi energy, are favored. 
In particular, these considerations seem to favor WSMs with only a few pairs of Weyl nodes.  

Several candidate TRS breaking WSMs have been recently synthesized
~\cite{sushkov,borisenko}; numerous other candidates 
have been proposed~\cite{wan,yingran,balents,tong,bulmash}. Most recently, magnetic Heusler alloys have been predicted to host only two Weyl nodes at the Fermi energy~\cite{bernevig16}. 

The twin hyperbolic and chiral character of FAPs may yield new photonic tools such as an intrinsically non-reciprocal Purcell enhancement of spontaneous emission 
for emitters placed close to a WSM surface. 
Along with the prospects of using FAPs as dynamical probes of the peculiar features of WSMs, these opportunities make the search for optimal materials for realizing FAPs, and detailed material-specific modeling of FAP characteristics, important directions for future exploration.

\vspace{2mm}
{\bf Acknowledgements ---} We thank Y. D. Chong, K. Flensberg, and M. S. Schecter for helpful comments on the manuscript. This work was supported by the Singapore National Research Foundation (NRF) under NRF fellowship award NRF-NRFF2016-05 (J.C.W.S.), and by the Villum Foundation (M.S.R).

%\newpage

\section{Supplementary Information}

\subsection{Vanishing bulk metallic contribution to surface velocity}
In an ordinary metal with no topological surface states, an excess surface density $n_s^{M} (\omega)$ 
may accumulate when the metal is pushed out of equilibrium. 
This surface charge layer may host a surface current density $e\vec v_{s}^M = \boldsymbol{\sigma}_{s}^M (-\nabla \phi)$, where $\boldsymbol{\sigma}_{s}^M$ is the surface conductivity associated with the 
bulk metallic carriers accumulated on the surface. 
Taking a simple Drude model for the longitudinal surface conductivity, we write $\sigma_{s}^M = n_s^M e^2/(m i\omega)$.  
Here $m$ is the effective mass of the carriers. 
Writing $n_s^M = n_s^{M,(0)} +  \delta n_s^M$ and noting that, in equilibrium, $n_s^{M,(0)} =0$, we find $\sigma_{s}^M \propto \delta n_s^M$. 
Similarly, because $\phi(\vec r, \omega) =  e\int U(\vec r, \vec r') \delta n(\vec r', \omega) d\vec r'$ is generated by the plasmon's density inhomogeneity, it also depends directly on $\delta n$. 
(Here $U(\vec{r},\vec{r}')$ is the Coulomb kernel.)
As a result, the surface velocity $\vec v_s^M$ supported by bulk carriers accumulated at the surface  
goes as $(\delta n)^2$. 
Hence, to {\it linear order} in $\delta n$, the surface current $\vec{v}_s$, Eq.~(4) of the main text, is supported solely by the Fermi arc surface state carriers.

%\vspace{2mm}
\subsection{Velocity density of topological surface states}

\subsubsection{A. Chiral velocity sign, and bulk-edge correspondence}
Here we resolve the sign of the chiral velocity $v_s^0$, Eq.~(\ref{eq:vel}) of the main text, associated with the topological surface states.
Consider a potential $\mathcal{V}$ that confines WSM carriers to the region $z<0$. 
This confining potential produces a force in the $-\hat{\vec{z}}$ direction, so that $e(- \partial_z \mathcal{V}) \propto - \hat{\vec{z}} $.  
Next we note that besides confining the carriers, $\mathcal{V}$ may also give rise to an anomalous Hall current $\sigma_{xz} (-\partial_z \mathcal{V})$. 
Due to the bulk-edge correspondence, the undergap anomalous Hall currents and edge currents from topological surface states move in the same direction. 
As a result, the surface velocity density satisfies
\be
{\rm sgn}(v_s^0) = {\rm sgn} (-\sigma_{xz}/e^2) = {\rm sgn}(\sigma_H).
\label{eq:signvel}
\ee
Why is there a minus sign in the middle expression? The confining electric field points towards $-(1/e)\hat{\vec{z}}$, giving an electric current in the direction ${\rm sgn}(-\sigma_{xz}/e)\hat{\vec{x}}$. %;
Further, the velocity and electrical current are related by another factor of $e$, giving the factor of $1/e^2$. 
Hence the sign of $v_s^0$ is {\it independent} of the sign of the carrier charge, $e$.
In the last expression we used $\sigma_H = \sigma_{zx} = - \sigma_{xz}$. 
Note that on the opposite surface the confining force would point in the $+\hat{\vec{z}}$ direction, yielding an opposite sign for the chiral velocity on that surface: ${\rm sgn}(v_s^{0, \rm opp}) = {\rm sgn}(-\sigma_H)$. 

\subsubsection{B. Drude model for Fermi arc $\hat{\vec{y}}$-direction velocity density}
The topological surface states of the WSM are characterized by a chiral velocity, oriented in the $\hat{\vec{x}}$ direction (i.e., the direction perpendicular to the $k$-space line connecting the bulk Weyl nodes). 
Additionally, for a generic topological surface state dispersion, a non-equilibrium Fermi arc carrier distribution may carry a velocity density in the $\hat{\vec{y}}$-direction, {\it parallel} to the line connecting the bulk Weyl nodes.
The precise features of these $y$-currents depend 
on the details of the topological surface state dispersion, and the scattering mechanisms on the surface. 
However, we note that the carriers may move in both the $+\hat{\vec{y}}$ and $-\hat{\vec{y}}$ directions, and at equilibrium (for zero electric field), the velocity density in the $\hat{\vec{y}}$-direction vanishes, $v_{s,y}^{(0)} =0$. 
Hence, 
we use a simple phenomenological Drude model to capture the $y$-currents induced by in-plane electric fields: 
\be
e v_y^s = \sigma_{yy}^s E_y = \frac{\mathcal{D}}{i\omega + \gamma} E_y, 
\ee
where $E_y = -\partial_y \phi$ is the electric field in the $\hat{\vec{y}}$ direction, and $\gamma$ is the transport scattering rate along the surface. 
In the collisionless limit, $\omega \gg \gamma$, the conductivity reduces to $ \sigma_{yy}^s \to \mathcal{D}/i \omega$. 

We note that, unlike the response to $y$-fields described above, 
the chiral surface states do not possess a Drude type response to $x$-directed electric fields. 
While the application of an electric field may impart momentum to the carriers along $\hat{\vec{x}}$, the linear, chiral dispersion along $\hat{\vec{x}}$ ensures that, to leading order (at fixed density), there is no change to the $x$-component of velocity density. 
However, as described in the main text, an $x$-directed electric field causes the topological surface state density to change due to impinging currents brought about via the bulk Hall conductivity.
Thus the velocity density $\vec{v}_s$ on the surface responds to electric fields in the $\hat{\vec{x}}$ and $\hat{\vec{y}}$ directions very differently, as captured by Eq.~(\ref{eq:vel}) of the main text.

%\vspace{2mm}
\subsection{FAPs in WSMs with multiple Weyl node pairs}
Fermi arcs also exist in WSMs with multiple bulk Weyl node pairs, labeled by an index $i =1, 2, ... $. 
Extending the two-fluid model used in the main text, we can associate bulk and surface densities $n_{b,i}$ and  $n_{s,i}^\chi$ with each (where $\chi = \{M, F\}$ labels the bulk free carrier and topological surface state contributions to the surface density, respectively). In the same limit $\omega\tau_{\rm surf} \gg 1$ considered in the main text, the fields $n_{s,i}^F$ and $n_{s,i}^M$ obey distinct equations of motion, similar to Eq.~(\ref{eq:eom}). Surface plasmons arising from the collective dynamics of these carrier densities can be obtained in the same fashion as detailed in the main text, with all components coupled through the common electrical potential $\phi$. For simplicity, throughout this section we take $\kappa = 1$.

To illustrate this approach, we confine ourselves to a system with two pairs of Weyl nodes, $i = 1,2$.
We take all four Weyl nodes to be situated in the $k_x$-$k_y$ plane, at $\pm \vec d_i^\parallel$ (i.e., $\vec d_i^\parallel \cdot \hat{\vec{z}} = 0$). 
Each pair of Weyl nodes contributes a bulk Hall conductivity $\sigma_{H,i}$ with sign (and orientation) consistent with that described in Eq.~(\ref{eq:signvel});
we will assume that the longitudinal conductivity $\tilde{\sigma}_{xx}$ is the same for all Weyl nodes. 
Here $\tilde{\sigma}_{xx}$ is the contribution to the total longitudinal conductivity coming from a single Weyl node pair, $i$.
Finally, the associated Fermi arcs for each pair of  Weyl nodes possess a surface velocity density given by
\be
\vec v_{s,i} (\vec{r}_s)= \Big[v_{s,i}^{(0)} n_{s,i}^F (\vec{r}_s)\Big]  \widehat{\vec{d}}_i^\perp  - \Big[ \frac{\mathcal{D}_0}{i \omega} (\widehat{\vec{d}}_i^\parallel \cdot \nabla)  \phi(\vec{r}_s)\Big]\widehat{\vec{d}}_i^\parallel.
\ee
Here $\widehat{\vec{d}}_i^\parallel$ and $\widehat{\vec{d}}_i^\perp$ are unit vectors (in the $x$-$y$ plane) that describe the directions parallel and perpendicular to the vector connecting the $i$-th pair of Weyl nodes, respectively, with $\widehat{\vec{d}}_i^\perp = \widehat{\vec{d}}_i^\parallel \times \hat{\vec{z}}$. 
$\mathcal{D}_0$ describes the Drude weight for a single Fermi-arc, $i$.

Similar to Eq.~(\ref{eq:ns}) of the main text, the total accumulated surface density can be written in terms of the electric potential, $\delta \tilde{n}_s = \sum_{i,\chi} \mathcal{G}^{\chi}_i \tilde{\phi}^<$, with
\be
\mathcal{G}^M_i=  - \frac{q\tilde{\sigma}_{xx} }{ie \omega}, \quad  \mathcal{G}^F_i =  \frac{\sigma_{H,i} (\widehat{\vec{d}}_i^\perp \cdot \vec{q}) +\tfrac{\mathcal{D}_0}{\omega} (\widehat{\vec{d}}_i^\parallel \cdot \vec{q})^2 }{e(\omega  - v_{s,i}^{(0)} \widehat{\vec{d}}_i^\perp  \cdot \vec q)}. 
\label{eq:ns2}
\ee  

The collective modes (surface plasmons) are found by seeking the combinations of $\omega$ and $\vec{q}$ such that the boundary conditions of continuous $\tilde\phi_{\vec{q}}$ and of the jump in the electric displacement field can be satisfied, as described above Eq.~(\ref{eq:decoupledsecular}) in the main text. 
Thus we must solve the generalized secular equation 
\be
- 2q + 4\pi e\sum_{i,\chi} \mathcal{G}^{\chi}_i = 0.
\ee
To derive the dispersion relation, analogous to Eq.~(\ref{eq:decoupledsecular}), we first write $ 4\pi\tilde{\sigma}_{xx}/\kappa =  \omega_{\rm pl}^2/(2i\omega)$, where $\omega_{\rm pl}$ is the bulk plasmon frequency of the WSM.
Here, the factor of 2 in the denominator on the right hand side reflects the fact that, in this example, the two Weyl node pairs each contribute half the total density of bulk free carriers.
Collecting terms with factors $(\omega - v_{s,i}^{(0)} \widehat{\vec{d}}_i^\perp  \cdot \vec q)$ and $(2- \omega_{\rm pl}^2/\omega^2)$, we obtain: 
\begin{widetext}
\begin{align}
 2(\omega - v_{s,1}^{(0)} \widehat{\vec{d}}_1^\perp  \cdot \vec q) \, (\omega - v_{s,2}^{(0)} \widehat{\vec{d}}_2^\perp  \cdot \vec q)\, \Big[1- \frac{(\omega_{\rm pl}^{\rm surf})^2}{\omega^2}\Big]  
 + \mathcal{M}_1 + \mathcal{M}_2 =0,
\label{eq:general}
\end{align}
where $\omega_{\rm pl}^{\rm surf} = \omega_{\rm pl}/\sqrt{2}$ (taking $\kappa =1$), and the hybridization amplitudes are given by
\bea
\nonumber \mathcal{M}_1 &=&  (\omega - v_{s,2}^{(0)} \widehat{\vec{d}}_2^\perp  \cdot \vec q) \Big[- 4\pi \sigma_{H,1} (\hat{\vec{q}} \cdot \widehat{\vec{d}}_1^\perp) - \frac{4\pi \mathcal{D}_0|\vec q|}{\omega} (\hat{\vec{q}} \cdot \widehat{\vec{d}}_1^\parallel )^2\Big],\\
\mathcal{M}_2 &=&  (\omega - v_{s,1}^{(0)} \widehat{\vec{d}}_1^\perp  \cdot \vec q) \Big[-4\pi \sigma_{H,2} (\hat{\vec{q}} \cdot \widehat{\vec{d}}_2^\perp) - \frac{4\pi \mathcal{D}_0|\vec q|}{\omega} (\hat{\vec{q}} \cdot \widehat{\vec{d}}_2^\parallel )^2\Big].
\eea
We note that for non-vanishing $\mathcal{M}_{1,2}$, the collective oscillations in the chiral branches $(\omega - v_{s,i}^{(0)} \widehat{\vec{d}}_i^\perp  \cdot \vec q) = 0$ and the conventional surface plasmon mode described by $(1- ({\omega}_{\rm pl}^{\rm surf})^2/\omega^2) = 0$ will hybridize, giving rise to FAPs. 
The structure of the resulting plasmon bands will depend on the magnitudes of the hybridization amplitudes, as well as the positions and orientations of the pairs of Weyl nodes (as encoded in the $\{\vec d_i\}$).

Seeking hyperbolicity, we note that, at large $q$, the terms going as $\mathcal{O}(q^2)$ dominate Eq.~(\ref{eq:general}).
Using this large $q$ limit, we determine the asymptotic contours via:
\begin{align}
 2  (v_{s,1}^{(0)} \widehat{\vec{d}}_1^\perp  \cdot \vec q) (v_{s,2}^{(0)} \widehat{\vec{d}}_2^\perp  \cdot \vec q)\Big[1- \frac{(\omega_{\rm pl}^{\rm surf})^2}{\omega^2}\Big] = - \frac{4\pi \mathcal{D}_0 |\vec q|}{\omega} \big[ v_{s,2}^{(0)} \widehat{\vec{d}}_2^\perp  \cdot \vec q (\hat{\vec{q}} \cdot \widehat{\vec{d}}_1^\parallel )^2 + v_{s,1}^{(0)} \widehat{\vec{d}}_1^\perp  \cdot \vec q (\hat{\vec{q}} \cdot \widehat{\vec{d}}_2^\parallel )^2  \big].
\label{eq:general-asymptotic}
\end{align}
\end{widetext}
As a sanity check, we note that for a WSM with two pairs of Weyl nodes that point in the same direction, $\vec d^\parallel_1 = \vec d^\parallel_2$ (and $v_{s,1}^{(0)} = v_{s,2}^{(0)} = v_0$), Eq.~(\ref{eq:general-asymptotic}) yields the same hyperbolae as described in the main text (setting $\kappa=1$ and recalling $\mathcal{D}_0 = \mathcal{D}/2$).

We now illustrate the contours of FAPs in a situation with multiple (2) pairs of Weyl nodes, with $\vec d_1^\parallel$ and $\vec{d}_2^\parallel$ pointing in different directions.
For demonstration, we take a simple model with $\vec{d}_1^\parallel$ along $\hat{\vec{x}}$ and $\vec{d}_2^\parallel$ along $\hat{\vec{y}}$:
\begin{align}
& \vec{d}_1^\parallel = \hat{\vec{x}}, \quad \quad \quad \quad \quad \quad \vec{d}_2^\parallel = \hat{\vec{y}}, \nonumber \\
& \vec{d}_1^\perp = \vec{d}_1^\parallel \times \hat{\vec{z}} = -\hat{\vec{y}}, \quad \vec{d}_2^\perp = \vec{d}_2^\parallel \times \hat{\vec{z}} = \hat{\vec{x}},
\label{eq:2pairs}
\end{align}
and $v_{s,1}^{(0)} = v_{s,2}^{(0)} = v$. 
Substituting the relations in Eq.~(\ref{eq:2pairs}) into Eq.~(\ref{eq:general-asymptotic}), we find  
open constant frequency contours that asymptotically approach the lines $\theta = \theta_\infty (\tilde{\omega})$ at large $q$, with
\be
\frac{{\rm cos}^3 \theta_\infty -{\rm sin}^3 \theta_\infty}{{\rm cos}\theta_\infty{\rm sin}\theta_\infty} = \frac{\tilde{\omega}}{\tilde{\mathcal{D}_0}} \Big[ 1- \frac{1}{\tilde{\omega}^2}\Big],
\ee
where $\tilde{\omega} = \omega/\omega_{\rm pl}^{\rm surf}$. 
Here $\tilde{\mathcal{D}_0} = 2\pi \mathcal{D}_0/(v \omega_{\rm pl}^{\rm surf})$, as in the main text (with $\kappa = 1$). 
A complex pattern of asymptotic contour lines $\theta = \theta_\infty(\tilde{\omega})$ arises from the hybridization of ordinary surface plasmons modes with collective oscillations in the Fermi arc surface states, and from hybridization between the collective modes of each of the separate branches of Fermi arc surface states (associated with the Weyl node pairs $i = 1,2$).


\begin{thebibliography}{99}
%first papers predicting Weyl semimetals [all in 2011]
\bibitem{wan} X. Wan, A. M. Turner, A. Vishwanath, S. Y. Savrasov, Topological semimetal and Fermi arc surface states in the electronic structure of pyrochlore iridates, Phys. Rev. B {\bf 83}, 205101 (2011).

\bibitem{balents} A. A Burkov, and L. Balents, Weyl semimetal in a topological insulator multilayer, Phys. Rev. Lett. {\bf 107} 127205 (2011).

\bibitem{yingran} K.-Y. Yang, Y.-M. Lu, Y. Ran, Quantum Hall effects in a Weyl semimetal: Possible application in pyrochlore iridates, Phys. Rev. B {\bf 84}, 075129  (2011). 

\bibitem{xugang} X. Gang {\it et al.}, Chern semimetal and the quantized anomalous Hall effect in HgCr$_2$Se$_4$, Phys. Rev Lett. {\bf 107}, 186806 (2011).

\bibitem{ding15} B. Q. Lv {\it et al.}, Experimental discovery of Weyl semimetal TaAs, Phys. Rev. X {\bf 5}, 031013 (2015).

\bibitem{hasan15} S.-Y. Xu {\it et al.}, Discovery of a Weyl fermion semimetal and topological Fermi arcs, Science {\bf 349}, 613-617 (2015).

%Review of WSMs
\bibitem{TurnerVishwanath}
A. Turner and A. Vishwanath, Beyond Band Insulators: Topology of Semi-metals and Interacting Phases, arXiv:1301.0330 (2013).

\bibitem{Haldane2004}
F. D. M. Haldane, Berry Curvature on the Fermi Surface: Anomalous Hall Effect as a Topological Fermi-Liquid Property, Phys. Rev. Lett. {\bf 93}, 206602 (2004).

%Fermi surface plumbing
\bibitem{HaldanePlumbing}
F. D. M. Haldane, Attachment of Surface ``Fermi Arcs'' to the Bulk Fermi Surface: ``Fermi-Level Plumbing'' in Topological Metals, arXiv:1401.0529 (2014).
\bibitem{sonspivak} D. T. Son, B. Spivak, Chiral anomaly and classical negative magnetoresistance of Weyl metals, Phys. Rev. B {\bf 88}, 104412 (2013).
\bibitem{Sid} S. A. Parameswaran, T. Grover, D. A. Abanin, D. A. Pesin, and A. Vishwanath, Probing the Chiral Anomaly with Nonlocal Transport in Three-Dimensional Topological Semimetals, Phys. Rev. X {\bf 4}, 031035 (2014).
\bibitem{potter} A. C. Potter, I. Kimchi, A. Vishwanath. Quantum oscillations from surface Fermi arcs in Weyl and Dirac semimetals, Nature Comm. {\bf 5}, (2014).

\bibitem{ong} J. Xiong, {\it et al.}, Evidence for the chiral anomaly in the Dirac semimetal Na3Bi, Science {\bf 350}, 413-416 (2015).
\bibitem{xidai} X. Huang, {\it et al.}, Observation of the chiral-anomaly-induced negative magnetoresistance in 3D Weyl semimetal TaAs, Phys. Rev. X {\bf 5}, 031023 (2015).
\bibitem{analytis} P. J. W. Moll, {\it et al.}, Transport evidence for Fermi-arc-mediated chirality transfer in the Dirac semimetal Cd3As2, Nature (2016).
\bibitem{spivakandreev} A. V. Andreev, B. Spivak, Magnetotransport phenomena related to the chiral anomaly in Weyl semimetals, Phys. Rev. B {\bf 93}, 085107 (2016).

\bibitem{baum} Y. Baum, et al,. Current at a distance and resonant transparency in Weyl semimetals, Phys. Rev. X {\bf 5}, 041046 (2015).

%paper by burkov explicitly discussed AHE in Weyl semimetals. While know before, i think he made the clearest discussion. 
\bibitem{burkov} A. A. Burkov, Anomalous Hall effect in Weyl metals, Phys. Rev. Lett. {\bf 113}, 187202 (2014).




\bibitem{son} D. T. Son, N. Yamamoto, Berry curvature, triangle anomalies, and the chiral magnetic effect in Fermi liquids, Phys. Rev. Lett. {\bf 109}, 181602 (2012).
% first surface plasmon paper by ritchie (after the famous bohm and pines paper that first talked about plasmons) ritchie showed they missed something in the 3D calculation.


\bibitem{sushkov} A. B. Sushkov {\it et al.}, Optical evidence for a Weyl semimetal state in pyrochlore Eu$_2$ Ir$_2$O$_7$, Phys. Rev. B {\bf 92}, 241108 (2015).

\bibitem{borisenko} S. Borisenko {\it et al.}, Time-reversal symmetry breaking type II Weyl state in YbMnBi$_2$, arXiv: 1507.04847 (2015). 

\bibitem{tong} T. Guan {\it et al.}, Evidence for half-metallicity in n-type HgCr$_2$Se$_4$, Phys. Rev. Lett. {\bf 115}, 087002 (2015).

\bibitem{bulmash} D. Bulmash, C.-X. Liu, X.-L. Qi, Prediction of a Weyl semimetal in Hg$_{1-x-y}$Cd$_x$Mn$_y$Te, Phys. Rev. B {\bf 89} 081106 (2014).

\bibitem{bernevig16} Z. Wang {\it et al.}, Time-Reversal-Breaking Weyl Fermions in Magnetic Heusler Alloys, Phys. Rev. Lett. 1{\bf 17}, 236401 (2016).

\bibitem{hyperbolicrev} A. Poddubny {\it et al.}, Hyperbolic metamaterials, Nature Photonics {\bf 7}, 948-957 (2013).

\bibitem{hoffman16} J. Hofmann, S. Das Sarma, Surface plasmon polaritons in topological Weyl semimetals, Phys. Rev. B {\bf 93}, 241402 (2016).

\bibitem{pesin} Panfilov, I., A. A. Burkov, D. A. Pesin, Density response in Weyl metals, Phys. Rev. B {\bf 89}, 245103 (2014). 

\bibitem{xiao} J. Zhou, H.-R. Chang, D. Xiao, Plasmon mode as a detection of the chiral anomaly in Weyl semimetals, Phys. Rev. B {\bf 91} 035114 (2015). 

\bibitem{Rosenstein} B. Rosenstein, H. C. Kao, and M. Lewkowicz, Nonlocal electrodynamics in Weyl semimetals, arXiv:1508.01604 (2015).

\bibitem{Cortijo} Y. Ferreiros and A. Cortijo, Unconventional electromagnetic mode in neutral Weyl semimetals, Phys. Rev. B {\bf 93}, 195154 (2016).

\bibitem{Lozovik} O. V. Kotov and Yu. E. Lozovi, Dielectric response and novel electromagnetic modes in three-dimensional Dirac semimetal films, Phys. Rev. B {\bf 93}, 235417 (2016).

\bibitem{SI} See {\bf Supplementary Information} for a discussion of Weyl semimetals with multiple pairs of Weyl nodes, vanishing bulk metallic contribution to surface velocity $\vec v_s^M$, and surface velocity density in the topological surface states.

\bibitem{fetter-b} A. Fetter, Edge magnetoplasmons in a two-dimensional electron fluid confined to a half-plane. Phys. Rev. B {\bf 33}, 3717 (1986). 


\bibitem{ritchie} R. H. Ritchie, Plasma losses by fast electrons in thin films, Phys. Rev. {\bf 106}, 874 (1957).


\bibitem{zyuzin} A. A. Zyuzin, V. A. Zyuzin, Chiral electromagnetic waves in Weyl semimetals, Phys. Rev. B {\bf 92}, 115310 (2015).
\bibitem{song16} J. C. W. Song, M. S. Rudner, Chiral plasmons without magnetic field, Proc. Natl. Ac. Sci. {\bf 113}, 4658-4663 (2016).


\bibitem{koppens2012} J. Chen {\it et al.}, Optical nano-imaging of gate-tunable graphene plasmons, Nature {\bf 487}, 77-81 (2012).
\bibitem{basov2012} Z. Fei {\it et al.}, Gate-tuning of graphene plasmons revealed by infrared nano-imaging, Nature {\bf 487}, 82-85 (2012).
\bibitem{alu} J. S. Gomez-Diaz, M. Tymchenko, A. Alu, Hyperbolic plasmons and topological transitions over uniaxial metasurfaces, Phys Rev. Lett. {\bf 114}, 233901 (2015).
\bibitem{low} A. Nemilentsau, T. Low, G. Hanson, Anisotropic 2D materials for tunable hyperbolic plasmonics, Phys. Rev. Lett. {\bf 116}, 066804 (2016).

\bibitem{parameters} In order to illustrate the properties of Fermi arc plasmons, we have used parameters $\tilde\sigma_H=2.0$, and $\tilde{\mathcal{D}} =6.0$. 
In arriving at these ball park values, we used $\kappa \sim 10$~\cite{sushkov}, and $\sigma_H = e^2 (2k_*) /h$ with $2k_* \sim 0.02 - 0.1 \, A^{-1}$. We also adopted a simple model for $\mathcal{D} \sim \bar{n}e^2/m$, where $\bar{n}$ is a typical surface density between the Weyl nodes; we used $m \sim E_F/\bar{v}^2$, with $E_F$ the Fermi energy. Taking $\omega_{\rm pl}^{\rm surf} \sim 20-60 \, {\rm meV}$, and a typical $y$-velocity on the surface $\bar{v} \sim 10^8 {\rm cm} \, {\rm s}^{-1}$, we arrive at order of magnitude estimates of $\tilde\sigma_H \sim 2 - 15$, and $\tilde{D} \sim 5 - 50$. 

\end{thebibliography}
\end{document}